# Two Types of Quaking and Shear Unjamming: State Diagram for Soft Granular Particles under Shear

Cheng-En Tsai (3,1), Wei-Chih Li 1, H.-C. Fan-Chiang 1, Pai-Yi Hsiao (2,*), and Jih-Chiang(JC) Tsai (1,*)
1) Institute of Physics, Academia Sinica, Taipei, Taiwan;
2) Department of Engineering and System Science, National Tsing Hua University, Hsinchu, Taiwan
3) Department of Phyiscs, National Taiwan University, Taipei, Taiwan
Correspondence (*): JC Tsai<jctsai@phys.sinica.edu.tw>

Understanding intermittency, an ubiquitous behavior in flows of packed grains, is pivotal for establishing the rheology of granular matter. A straightforward explanation has been missing despite the long development of theories at various levels of abstraction. Here, we propose the use of a Stribeck-Hertz model that starts with the classic Coulomb friction but takes into account the inter-particle *tribology*, *i.e.* the reduction of friction coefficient with sliding speed as is commonly observed. Our numerical experiments reveal a state diagram that covers a wide range of packing fractions and show that incorporating the tribology enables the occurrence of quaking intermittency in the mid-range of a newly established dimensionless shear rate, in consistence with prior experimental observations. Further study of the discontinuities in the evolution of mean contact number leads to our discovery of two types of quaking, that are distinguished by the abrupt *increase* or *decrease* of neighboring contacts and reveal different pathways of microstructural change underlying these discrete events. In contrast to the prevailing paradigm in which shear is believed to promote jamming at intermediate densities, our study demonstrates that shear can also unjam a granular system, and this occurrence depends on the shear rate.

## I. INTRODUCTION

Intermittency is an important signature in the flow of packed grains. It distinguishes granular flow from ordinary fluid and offers a perfect test ground for theories on rheology of such intriguing materials. Many phenomena, such as landslides and earthquakes, can be attributed to the occurrence of intermittency. Despite its significance, there have been relatively few studies dedicated to understanding its origin, in comparison to decades of extensive research on the rapid or creeping regime of granular flows [1–4] where intermittency is negligible. Over the years, various experimental techniques have laid the groundwork for understanding the force distribution [5], internal structure [6–10], and fluctuations [11,12] inside a steadily sheared granular pack. Recently, we conducted a laboratory experiment on granular shear flow which exhibited a strong intermittency, characterized by bursts of grain-level displacements and sudden releases of stress [13]. Interestingly, such intermittency occurred only *within an intermediate range of the driving rates*. The experiment also demonstrated a transition from the realm of “granular suspension” [14,15] to the state of “packed grains” where the elastic interaction among soft particles prevails over the drag exerted by the interstitial fluid.

Here, we perform an independent numerical study that not only reproduces the rate-dependent intermittency as reported in Ref. [13] but also leads to intriguing predictions going beyond what is available from existing laboratory observations. In retrospect, it has been well known that interparticle friction plays a significant role on the behaviors of a granular packing. Several milestone numerical works have shown that many properties such as the critical density for jamming [16,17], the mechanical stability for a static packing [18], and the reversibility in response to quasi-static driving [19] are quite *sensitive* to the strength of friction. However, in all preceding analyses, the “friction coefficient” is usually treated as a *constant*, known as the Coulomb friction. In reality, friction coefficients can vary with the state of their relative motion. How a granular system would behave with non-constant friction remains an open question. Recent studies using discrete modelling based on “load-activated friction”[23,24] have successfully explained the dramatic shear thickening (ST) of dense suspensions, while an alternative model based on pure hydrodynamics has also proved to work indistinguishably well [22]. But note that these ST models target mainly on the regime of high shear rates where fluid drag is essential and the system is not tightly packed. The ST models are therefore inadequate for explaining the occurrence of intermittency in quasi-static flows, such as those described in Ref. [13] where grains are elastically packed. On the other hand, advanced continuum theories have been developed to tackle the issue of non-constant friction with packed grains, where a “field of friction coefficient” (nonlocal effect) have been proposed but with breakthroughs yet to be anticipated [23–25]. In geoscience, numerous studies [26–30] have revealed the phenomena of dynamic fault weakening, where macroscopic friction dramatically decreases during earthquake activities. This has inspired numerous laboratory experiments along the tradition of rock mechanics [31–33]. However, there is still a lack of consensus regarding how various particle-level activities contribute to global intermittency, specifically the nucleation of earthquakes.

The key aspect of our numerical studies is to incorporate *inter-particle* tribology, where the "friction coefficient" varies with the sliding speed between solid surfaces, and to explore its influence on the collective behaviors of granular particles. It should be noted that, despite the paradigmatic Stribeck curve [34,35] has become a common knowledge in optimizing the performance of journal bearings for more than a century, the role of inter-particle tribology on the dynamics of granular flow has not received as much attention in the research community until a recent experiment by Dijksman and coworkers [36], to the best of authors' knowledge. In what follows, we will demonstrate that incorporating inter-particle tribology is essential in producing the *collective, rate-dependent* intermittency, leading to new insights beyond existing observations, and adding a keystone to decades of scholarly discussions over "shear" and "jamming" of granular matter. We start with Section II, which shows that a straightforward modification to the classic Coulomb friction between particles renders the quasi-static system no longer rate-independent, resulting in a multitude of state transitions across a wide range of packing fractions and driving rates. Section III provides a comprehensive analysis of the regimes exhibiting strong intermittency, also known as *quaking*. We specifically focus on the discovery of two *types* of quaking that are distinguished by sudden *increases* or *decreases* in particle contacts. Section IV delves into the fluctuation of elastic energy, setting the stage for discussing the unjamming of granular packing. In Section V, we introduce a conceptual State Diagram that distinguishes our findings from the classic understanding, and elucidate how and why incorporating a realistic tribology gives rise to scenarios of "shear unjamming".

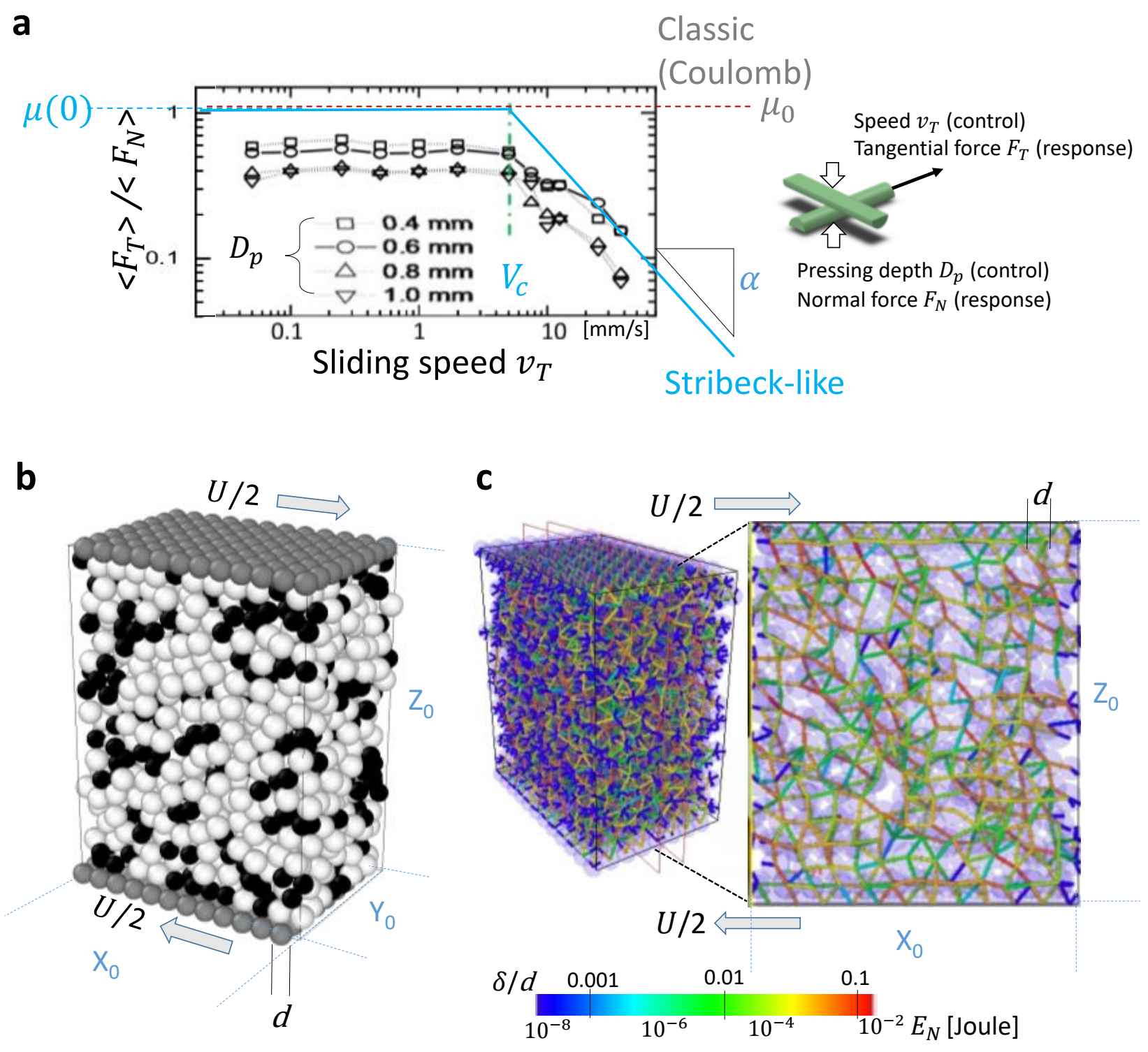

**FIG.1** -- Friction laws and visualization of our numerical studies -- **(a)** Stribeck-like friction coefficient $\mu(v_T)$, specified by the three parameters: the plateau value $\mu(0)$, the threshold speed $V_c$, and the reduction exponent $\alpha$, in comparison to the classic Coulomb friction specified by the single speed-independent parameter $\mu_0$. Data points show results of actual laboratory measurements [13,37,38]. **(b)** Snapshot of our numerical experiments for a granular system of two particle sizes ($0.9d$ shown in black and $1.1d$ in white). Shearing is effectuated by moving the top and bottom walls (made of the gray particles with diameter $d$) in opposing directions at a relative speed $U$. The separation of the two walls is $Z_0$, ranging from$14d$ to $20d$. The standard horizontal span of the simulation box is $X_0 = 12d$ by $Y_0 = 8d$ and periodic in both directions. **(c)** Visualization of individual contacts and the elastic energy stored between particle pairs, at some instant of shearing (with $\phi = 0.767$ and $S_\ell \sim 0.01$ under SH model). Every line connecting the centers of two particles represents the "depth" of contact ($\delta$) and the elastic energy ($E_N \propto \delta^{2.5}$), colored by the code shown as a bar. The panel on the right highlights a $1.4d$-thick slab along a vertical plane.

`

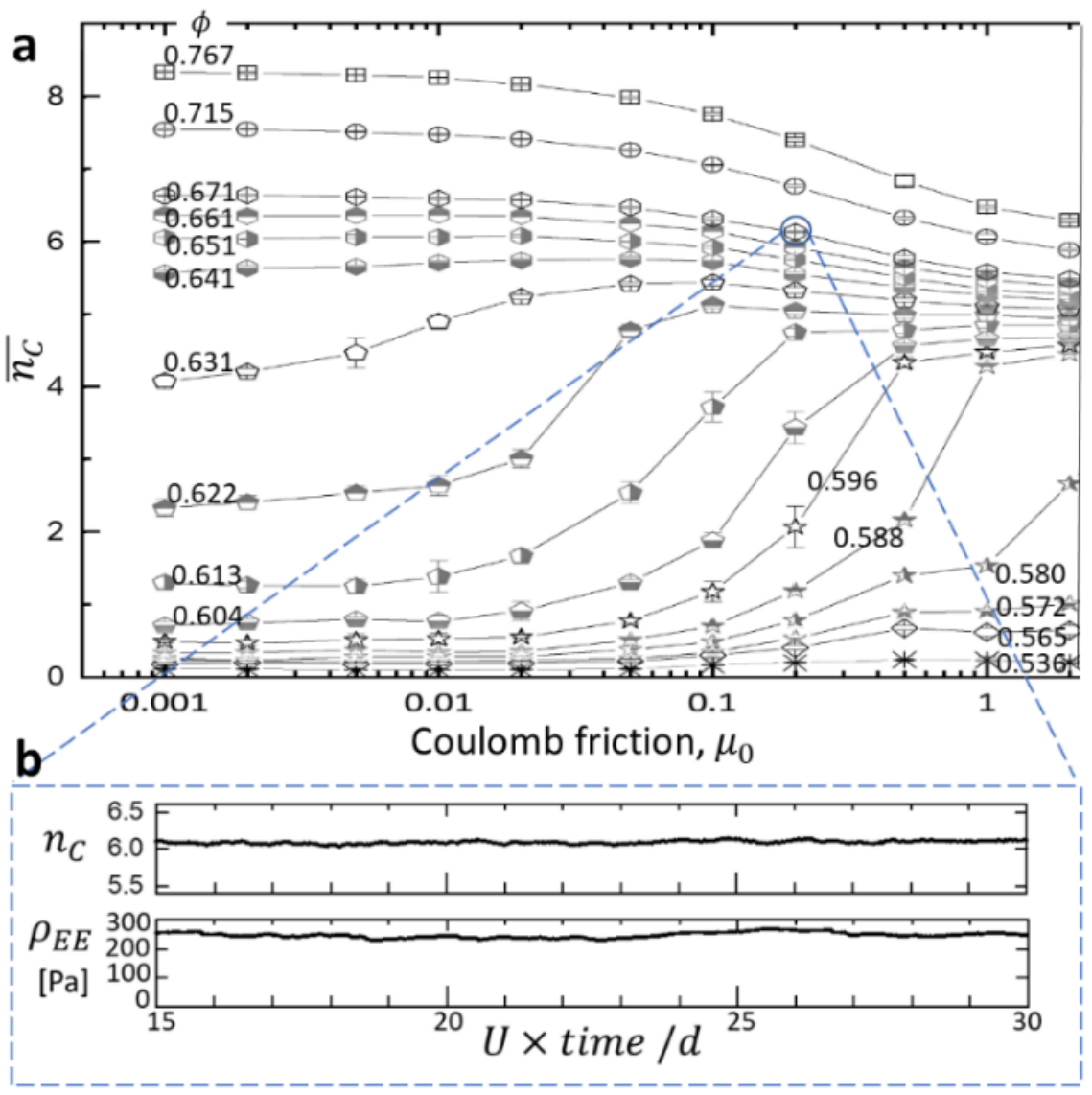

**FIG.2 --** Summary of results obtained by using the classic CH model at the low-inertia (quasi-static) limit. **(a)** Ensemble-averaged mean contact number (the coordination number), $\overline{n_C}$, as a function of the Coulomb coefficient $\mu_0$, for various packing fractions $\phi$. The bar around each data point represents the statistical error. **(b)** Time sequence of the instantaneous value of $n_C$ and the density of elastic energy ($\rho_{EE}$), showing no significant quaking intermittency. The case with $\phi = 0.671$ and $\mu_0 = 0.2$ is shown as one such example.

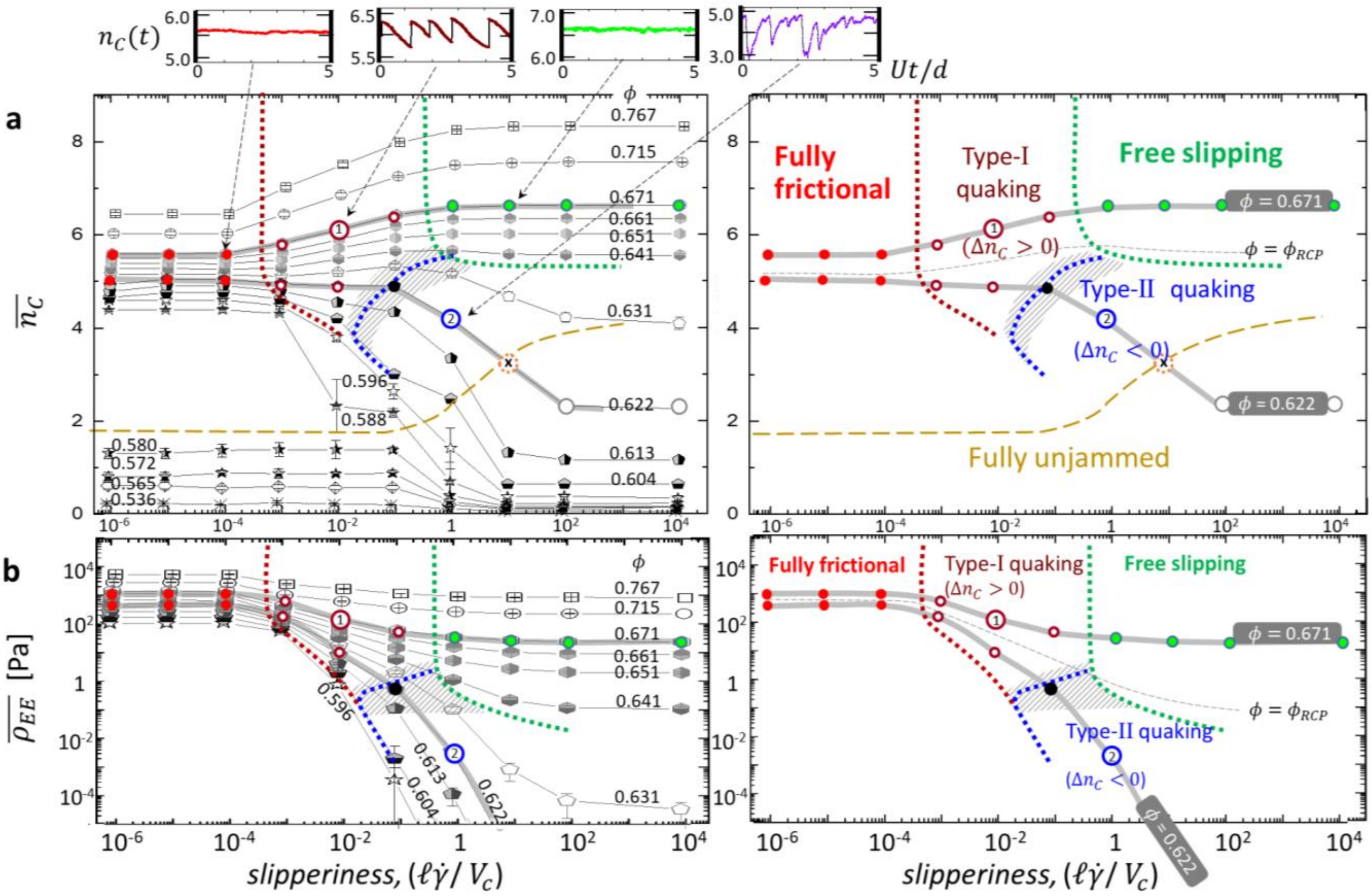

**FIG.3 --** Overview of the state transitions, based on simulations using SH model. **(a)** The ensemble-averaged value of $\overline{n_C}$ and **(b)** density of elastic energy $\overline{\rho_{EE}}$, as a function of the *slipperiness* $S_\ell$ for different values of packing fraction $\phi$. The two main panels on the left display data at 16 different values of $\phi$. Thick dotted lines represent boundaries of different regimes for the jammed packing. They are drawn based on the qualitative difference in the time sequence $n_C(t)$, of which examples are shown in the four mini-panels on top. The shaded area denotes the occurrence of a *mixed regime* where two types of quaking events coexist. The two panels on the right highlight two paths of constant packing fraction ($\phi = 0.671$ and $\phi = 0.622$) and the key properties of the dynamical regimes, to facilitate subsequent analyses. The thin dashed line (in gray) labeled with $\phi = \phi_{RCP}$ for the *random close packing* density [17] follows our data points at $\phi = 0.641$. The thick dashed line in the lower portion of (a) represents the boundary for "unjamming" – see main texts for further information.

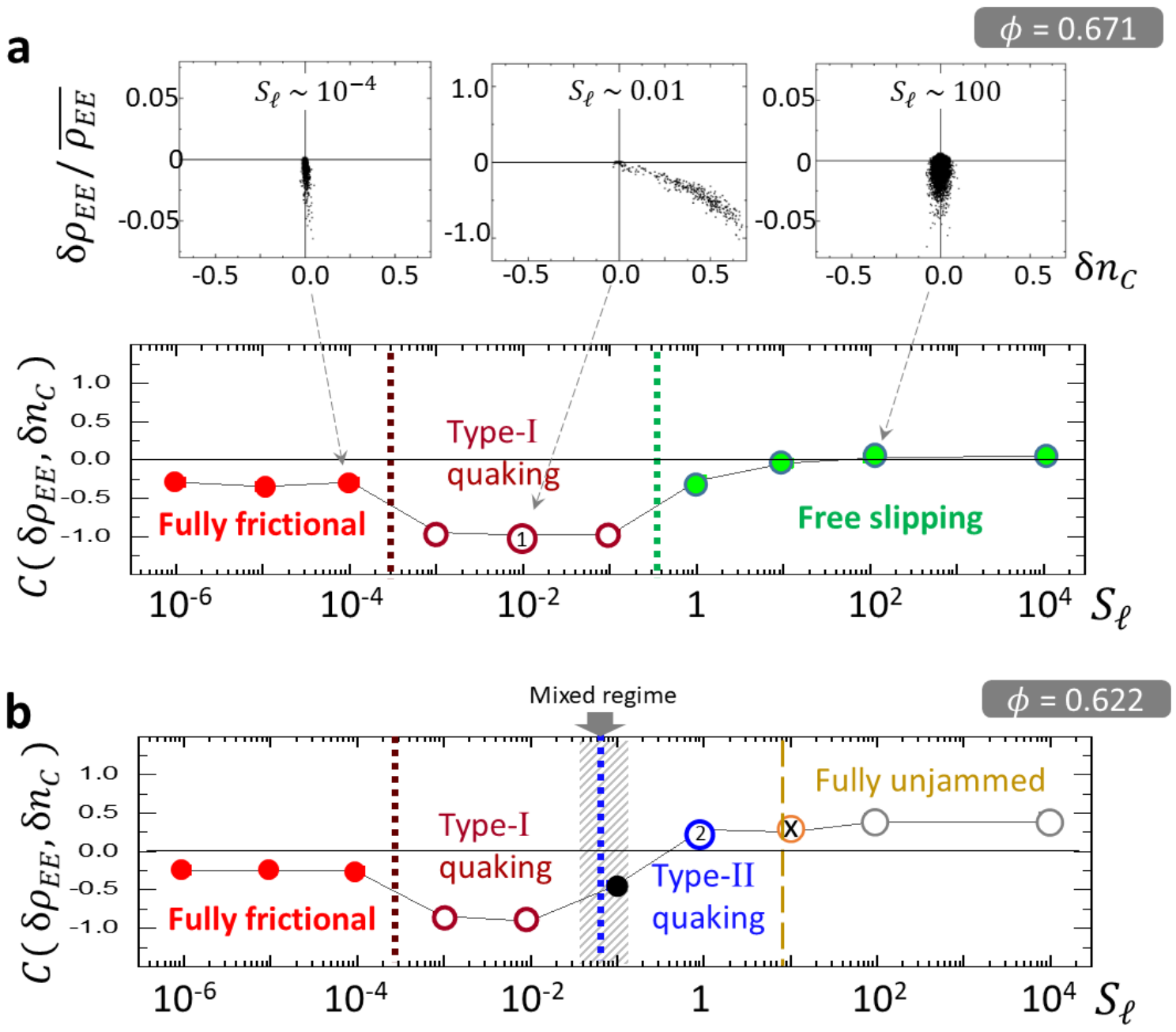

**FIG.4 --** Fluctuations observed along two representative paths of constant packing fraction. Dynamical regimes and their boundaries corresponding to those defined in Fig.3 are indicated with the same set of symbols. **(a)** Scatter plots of $\delta\rho_{EE}/\overline{\rho_{EE}}$ versus $\delta n_C$ at three different values of $S_\ell$, and the linear correlation coefficient $C(\delta\rho_{EE}, \delta n_C)$ as a function of $S_\ell$ along the path $\phi = 0.671$. **(b)** $C(\delta\rho_{EE}, \delta n_C)$ as a function of $S_\ell$ along the path $\phi = 0.622$.

## II. FRICTION LAWS, SHEAR FLOW, AND OVERVIEW OF STATE TRANSITIONS

Our modification to the classic friction law (the Coulomb friction) is specified by a speed-dependent friction coefficient, as shown in **Fig.1(a)**. It has a plateau value $\mu(0)$ up to a threshold speed $V_c$, followed by a power-law decay with an index of $\alpha$. The plateau and the threshold mimic the common Stribeck curve at its low-speed portion. This functional form has been verified independently in multiple works, including direct measurements using surfaces of polyacrylamide, gelatin [36], or polydimethylsiloxane (PDMS) [13,37,38]. We set $\alpha = 1$ as the default value for all simulations that follow, justified by the data cited in Ref.[13]. The data also show that $V_c$ is in general not sensitive to the normal load ($F_N$) or the pressing depth. On the other hand, the classic Coulomb friction, independent of the sliding speed, is shown as a dashed line and denoted by $\mu_0$. The geometry for shear flow is shown in **Fig.1(b)**: Densely packed soft spheres are sheared in a three-dimensional box of fixed volume, driven by two walls moving against each other at a speed $U$ at a fixed separation $Z_0$ and providing the shear rate, $\dot{\gamma} = U/Z_0$. Two sizes of particles with diameters $0.9d$ and $1.1d$, where $d$ denotes that of particles on the wall, are used to prevent the ordering of particles. The complete set of rules governing the force between contacting particles is referred as the Stribeck-Hertz (SH) model, in which the classic Hertzian law serves as the leading term for the normal force ---see **Appendix-A** for the full equations that define the contact force, the protocols in obtaining steady states with different initial configurations, and other implementation details. For comparison, counterpart simulations using a fixed friction coefficient, as has been widely used in prior studies [16,17,19], are denoted as using the Coulomb-Hertz (CH) model. The mean contact number ($n_C$) at each moment is defined as the instantaneous average of how many neighbors that a particle is in contact with, over the entire bulk, while its time averaged ($\overline{n_C}$) corresponds to the *coordination number* in literatures. We also compute the instantaneous elastic energy between particles from the governing equation, $E_N \propto K\,\delta^{2.5}$ where $K$ is the elastic modulus and $\delta$ is the depth of contact between particles. **Fig.1(c)** shows that the distribution of energy forms a highly anisotropic network against shearing at the particle level, which is reminiscent of prior studies on "laboratory fault" in 2D using photoelastic grains [39]. For the purpose of characterizing the state transitions, we use the spatial density of elastic energy ($\rho_{EE}$) and its fluctuation to represent what are obtained in boundary force measurements such as in Ref.[13], without computing the full stress tensor.

We first present in **Fig.2(a)** a state diagram that summarizes simulation results obtained from the classic picture (CH model), where the Coulomb coefficient is scanned over a wide range $\mu_0 = 0.001{\sim}2.0$. In the figure, the mean contact number ($\overline{n_C}$, or *coordination number*) is plotted as a function of $\mu_0$ for each specified value of *nominal packing fraction*, $\phi \equiv$ volume summed over all spheres / total volume of the box. Data points on the graph are the averages calculated from 10 independent initial configurations. We have verified (with data available as Fig.S1 in Supplemental Material) that the combination of our driving speed ($U$ = 0.1cm/s), the mass, the elastic modulus, and the packing fraction ensures that the particle dynamics lie within the low-inertial limit, commonly referred as the "quasi-static flow" for its force network is always very close to static equilibrium [16,17]. The density of kinetic energy (~ mass density$*\, d^2\dot{\gamma}^2$, given the strong dissipation among particles) is much smaller than the elastic energy density. It is noteworthy that, for the full range of packing fractions and Coulomb coefficients that we have explored, the fluctuations of these steady states are rather weak, illustrated in **Fig.2(b)**, with no signs of intermittency detected. In consistence with prior works [16,17], it is also well known that, within this low-inertia limit, the behavior of these states is "rate-independent" (insensitive to $U$) and is uniquely determined by the value of $\phi$ alone,

Implementing the SH model with its speed dependence in the friction coefficient $\mu$ leads to an important consequence: The state of the shear flow is no longer uniquely determined by the value of $\phi$, even if we still keep the system well within the low-inertia limit. Rather, the flow becomes *rate-dependent* as described by a dimensionless parameter that we call *slipperiness* or $S_\ell$. For generality, we define $S_\ell \equiv (\ell\dot{\gamma}/V_c)^\alpha$. It can be interpreted as a *factor* by which the average friction, in the presence of an imposed shear rate $\dot{\gamma}$, is *reduced* from its full effect at the slow limit ($\dot{\gamma} \to 0$). The value $\alpha$ is the power-law index that specifies the strength of the velocity weakening beyond the threshold speed $V_c$. Given the value $\alpha = 1$, the parameter $S_\ell$ is indeed a dimensionless shear rate for the flow. The mean particle diameter $d$ for our binary mixture is assigned as the characteristic length $\ell$, which can be generalized to other contexts by taking the distribution of particle sizes into account.

**Figure 3** shows a wealth of behaviors in SH simulations at various combinations of $S_\ell$ and $\phi$. For each packing fraction $\phi$, the value of $S_\ell$ spans over a wide range that is obtained by varying $V_c$ but keeping $U$ the same as that in producing Fig.2. This ensures that all states are within the low-inertia (quasi-static) limit, as is evidenced by Fig.S1 in Supplemental Material. One notable feature is that the variation of $\overline{n_C}$ in Fig.3(a) looks similar to the mirror image of the classic picture shown in Fig.2, because $S_\ell$ is the factor by which the friction coefficient between contacting particle pairs ($\mu_{ij}$, for instance), is reduced from that at its full value $\mu(0)$ as a global average, for the reason just described in defining $S_\ell$. Understandably, at high values of $S_\ell$, the shear flow is expected to exhibit free slipping behaviors, with the value of $\overline{n_C}$ approaching the zero-friction limit in Fig.2. In the other extreme, at low values of $S_\ell$, most particles behave as under full friction. Values of $\overline{n_C}$ therefore approach those at $\mu_0 = 1$ in Fig.2, because $\mu(0)$ has been set to $1$ in these SH simulations. Meanwhile, Fig.3(b) displays the time-averaged elastic-energy density $\overline{\rho_{EE}}$ as a function of $S_\ell$ for different packing fractions $\phi$, showing similar smooth changes that connect the free-slipping and the full-friction extremes.

The most salient feature of the results using SH model is the patterns of fluctuation in $n_C$ and $\rho_{EE}$, leading to multiple dynamical regimes that are displayed in Fig.3(a) and 3(b). In particular, for a substantial range of $\phi$ and $S_\ell$, the sheared packing exhibits *strong intermittency* that can be divided into two regimes: One dominated by "Type-I" quaking and the other by "Type-II", characterized by sudden *rises* and *drops* in the instantaneous value of $n_C(t)$, respectively. In Section III, we provide in-depth analyses on the two types of quaking. There is also a narrow crossover region where two types of quaking occur alternatively – details of this mixed regime is supplied as part of the supporting data (Fig.S3a) in Supplemental Material (SM). At low values of $\phi$ (and partially conditional on $S_\ell$), the granular packing can become *unjammed* – this will be revisited in Section IV and V with the support of statistical analyses on the time fluctuation of $\rho_{EE}$.

Along the two contours of constant packing fraction $\phi = 0.671$ and $\phi = 0.622$ as highlighted on the right panels of Fig.3(a) and 3(b), we further illustrate the state transition by correlating the fluctuations of $n_C$ and $\rho_{EE}$**.** The results are summarized in **Fig.4.** Here, the "instant changes" of contact number and elastic energy, $\delta n_C$ and $\delta\rho_{EE}$, are computed at a fixed time interval $\delta t$. Such $\delta t$ corresponds to a boundary displacement of $0.01d$ (or equivalently a strain of about $0.006$) that is small enough for identifying individual quaking events. We analyze the correlation between $\delta n_C$ and $\delta\rho_{EE}$ along the two paths of $\phi$ over the change of $S_\ell$. To illustrate the transitions along the path $\phi = 0.671$, Fig.4(a) shows the linear correlation coefficient $C(\delta\rho_{EE}, \delta n_C)$ as a function of $S_\ell$, with three representative scatter plots of $\delta\rho_{EE}/\overline{\rho_{EE}}$ versus $\delta n_C$ above the main graph. Note the pattern in the scatter plot for the quaking state, that explains how the Type-I quaking events contribute to the strong negative correlation between $\delta n_C$ and $\delta\rho_{EE}$. Fig.4(b) shows that, along the path $\phi = 0.622$, a similar negative correlation occurs for the states dominated by Type-I quaking, while this graph also shows a crossover of $C(\delta\rho_{EE}, \delta n_C)$ from negative to positive values as the system goes progressively through the mixed state, Type-II quaking, and eventually states with the packing fully unjammed --- see supporting data Fig.S3(b-d) in SM for further discussion. In addition, we provide evidence that the occurrence of strong quaking intermittency requires only a mild velocity weakening in the SH model at a small value of α, such as 0.25. The supporting data is available as Fig.S2 in SM.

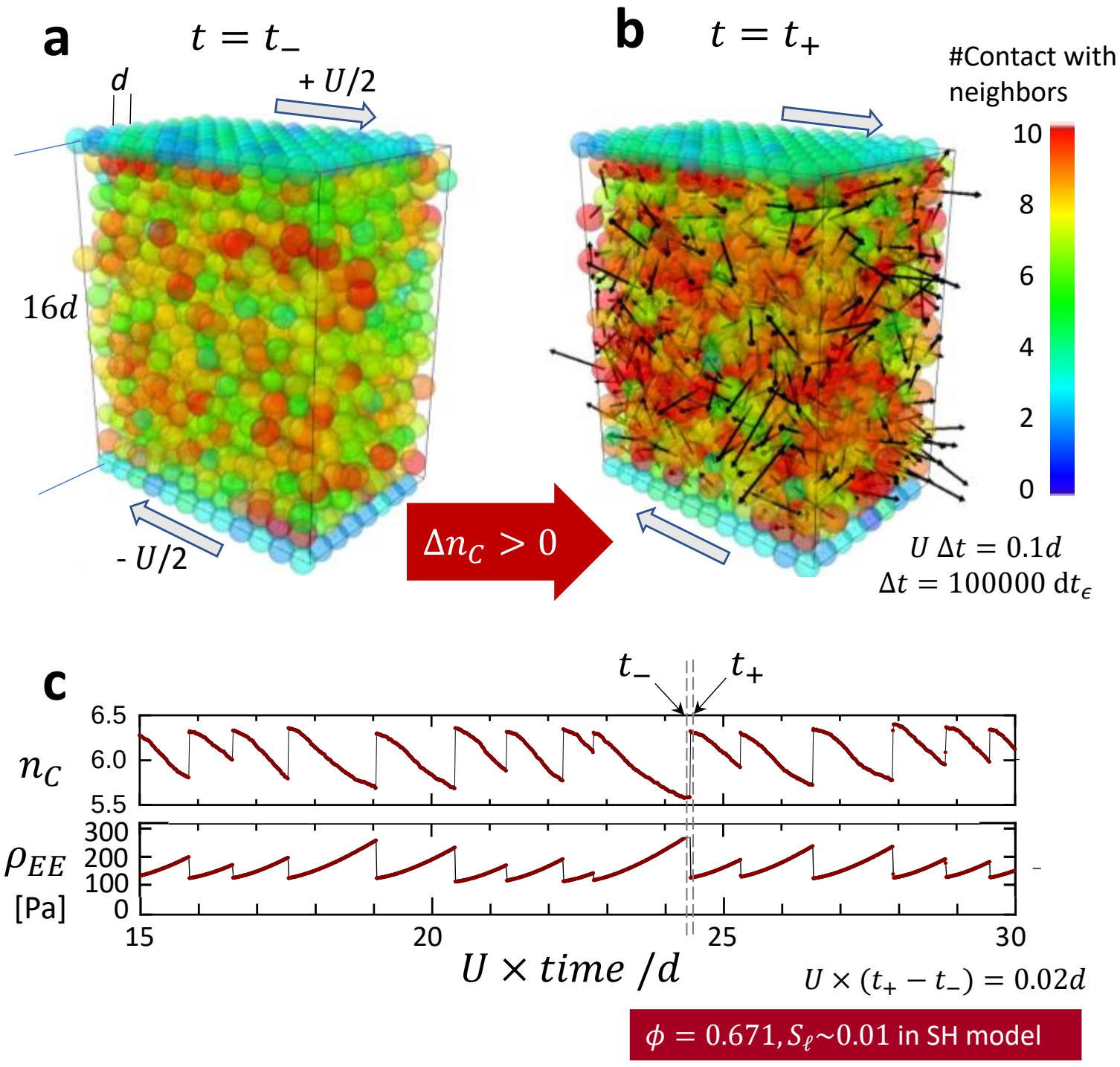

**FIG.5** Features of Type-I quaking ( $\phi = 0.671$ , $S_\ell \sim 0.01$, SH model). Number of contacts for each particle at the two instants **(a)** before ($t = t_-$) and **(b)** after ($t = t_+$) a quaking event, with the color code shown as a bar. The arrows plotted at the center of each particle indicate the displacements of particle from $t - \Delta t$ to $t$. In both panels, the length of the arrow has been amplified by a factor 10 for easy visualization. **(c)** Time sequence of $n_C$ and $\rho_{EE}$ in simulations. See also **Movie 1 online.**

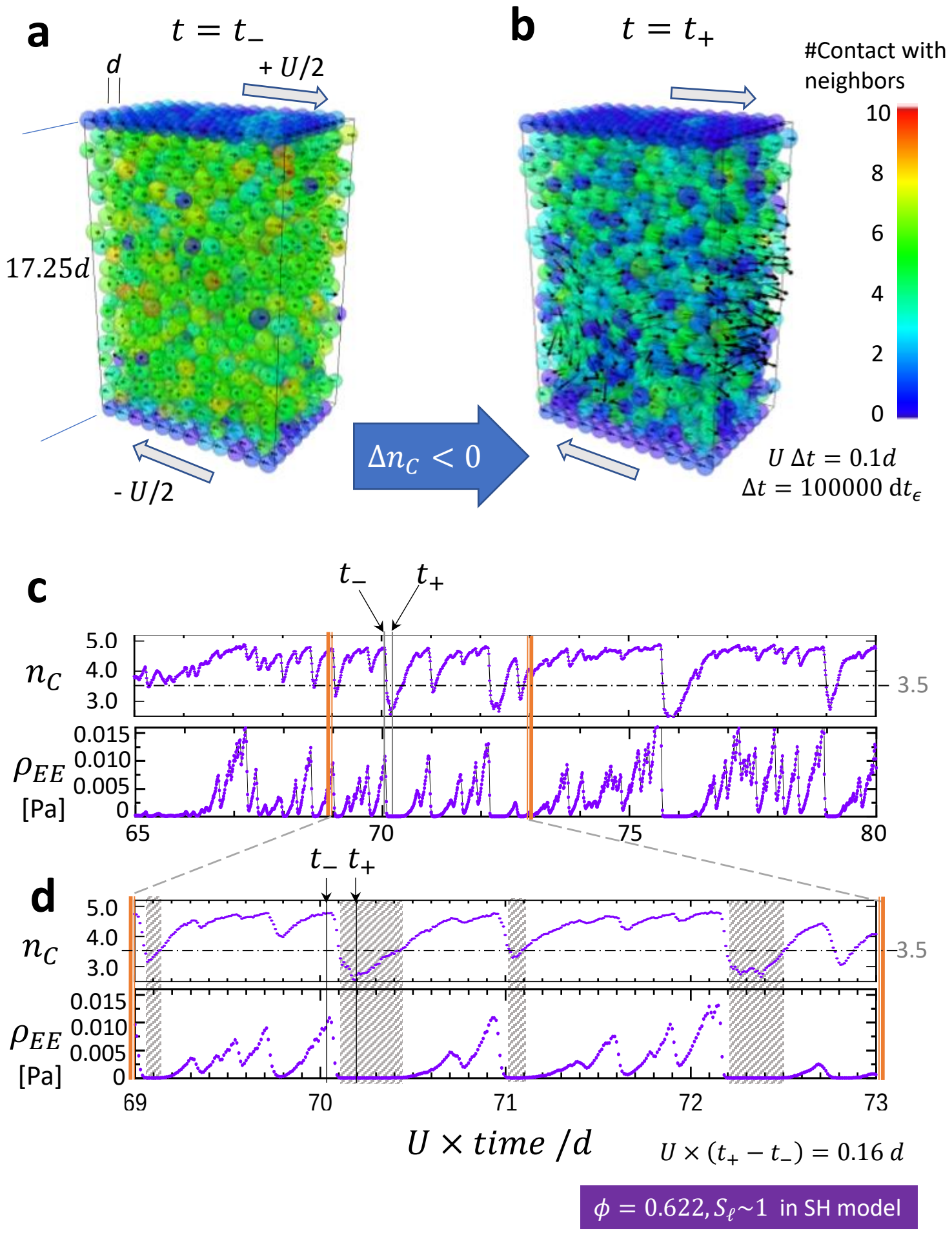

**FIG.6** Features of Type-II quaking ($\phi = 0.622$, $S_\ell \sim 1$, SH model), using the same color coding, definition of $t_-$ and $t_+$ and the magnification factor for displacements over $\Delta t$ as those in Fig.5. **(a, b)** Visualization of one event. In view of the displacement vectors, the contrast between $t_-$ and $t_+$ here is not as sharp as that for a Type-I event [Fig.5a-b]. **(c)** Time sequence of $n_C$ and $\rho_{EE}$ at the same horizontal scale as that in Fig.5. **(d)** Close-up plot of panel C with a shear strain of about 0.25. The intervals where the mean contact drops below 3.5 are marked with shades. See also **Movie 2 online**.

## III. TWO TYPES OF QUAKING AND EXPLANATIONS

At sufficiently high packing fractions $\phi > \phi_{RCP} \approx 0.64$, the transition from the state of full friction and that of free slipping is mediated by a regime of Type-I quaking, characterized by an abrupt rise of the instantaneous value of $n_C$ with each event. This is visualized in **Fig.5(a-b)** and with "$\Delta n_C > 0$" for short. In addition to the neighbor counting, displacements of individual particles (at a strain step $U\Delta t/Z_0 \approx 0.006$) have been indicated on the two graphs, for comparison. Frame-by-frame inspection of the **movie** (available in SM) further demonstrates that the movements of particles are far from uniform. The sudden displacements or "jumps" in Fig.5(b) can reach a significant fraction of the grain size, in contrast to the ones in Fig.5(a) in which the displacements are almost invisible at the same scale. The patterns of $n_C$ and $\rho_{EE}$ with repetitive occurrence of Type-I cycles are shown in **Fig.5(c).** Qualitatively speaking, the quaking events can be seen as a consequence of the distributed values of $\mu$ among particles, combined with the instability associated with the Stribeck-like speed dependence. A substantial amount of low-$\mu$ contacts provides an occasion for an "avalanche" (the quake) to be triggered. Such avalanche terminates when a significant fraction of particles is stopped by new neighbors. This explains the sudden increase of $n_C$ right after a quaking event. Beyond each event, the shearing makes incremental changes to the packing, and the newly established frictional contacts foster (a) anisotropic detachments of many "weak links" that are orthogonal to the "strong chains" [40] and (b) steady rise of elastic energy. This process continues until the occurrence of the next quake, thus explaining the relatively smooth decrease of $n_C$ between quaking events. In contrast to the classic CH model in which all particle contacts share the same friction coefficient ($\mu_0$), it is the speed-dependent values of $\mu$ in our SH model that underlies the instability for producing the quaking intermittency.

In the intermediate packing fractions between $\phi_{RCP}$ and a lower bound $\phi_0$ (that is determined by the value of $\mu(0)$ set in the SH model and can be estimated from observing Fig.2), we find that the dominant type of quaking depends partly on the value of $S_\ell$. At $\phi = 0.622$ and $S_\ell \sim 1$, for instance, we find that the instantaneous contact number $n_C$ can exhibit a drop during the quaking. We refer to these events as Type-II -- see **Fig.6** with the label "$\Delta n_C < 0$", for a side-by-side comparison to Fig.5. We explain such crossover from the dominance of Type-I events to that of Type-II from two complementary aspects. (a) From the aspect of mean contact number $\overline{n_C}$ (also known as the *coordination number*), the $\overline{n_C}$-$S_\ell$ diagram in Fig.3(a) has demonstrated that increasing the dimensionless shear rate to $S_\ell \sim 1$ along the path $\phi = 0.622$ reduces the value of $\overline{n_C}$ to be around 4. The study of isostaticity has established that the packing of frictionless spheres in a 3D space demands a coordination number 6 for its mechanical stability, and that incorporating particle frictions would lower this threshold value, which turns out to be 4 at the limit of an infinite friction coefficient [16–18]. Since $S_\ell \sim 1$ suggests that the packing is only partially frictional, the mechanical structure for such a state with $\overline{n_C} \approx 4$ is expected to be highly unstable, in the presence of Type-II quaking. In comparison, the $\overline{n_C}$-$S_\ell$ diagram has shown that the states dominated by Type-I quaking are either more frictional (with lower value of $S_\ell$) or have a higher number of contacting neighbors (with an average coordination number $\overline{n_C}$ close to 6 or higher) than those dominated by Type-II quaking. (b) The second aspect is obtained by observing the $\overline{\rho_{EE}}$-$S_\ell$ diagram in Fig.3(b) that allows us to estimate the typical depth of contact. Given that the energy density $\overline{\rho_{EE}}$ for a state around the borderline between the Type-I and Type-II regimes is around 1 Pa, the typical depth of contacts of such a state is approximately $10^{-3}d$, using the estimate $\overline{\rho_{EE}} \sim (\overline{n_C}/2)\, d^{0.5}\, K\, \delta^{5/2}/d^3$ based on a simple geometrical consideration. Below the borderline, the overlaps among spheres are much shallower than $10^{-3}d$ as $\delta \propto \overline{\rho_{EE}}^{2/5}$. Such scenario of shallow contacts naturally promotes the detachment of existing contacts during each "slip" (between the time $t_-$ and $t_+$). As the shearing continues, the particles proceed to reestablish the lost connections, resulting in a gradual rise in $n_C$ between successive Type-II quaking events.

In **Fig.6(d),** we illustrate further that Type-II events are very often associated with a short interval where the values of $\rho_{EE}$ are flat and effectively zero ($< 10^{-5}$ Pa), despite that $n_C$ keeps increasing. This indicates that, in the presence of Type-II quaking, there is a measurable fraction of time in which the elastic network is temporarily disintegrated. This has an important implication for the unjamming of the granular packing, and will be revisited in Section V.

# IV. STATISTICS ON ENERGY FLUCTUATIONS AND THE "FULLY UNJAMMED" STATE

For all states of flow that are tightly *packed*, the time statistics of $\delta\rho_{EE}$ reveals a biased distribution, where the system spends less time in releasing the elastic energy ($\delta\rho_{EE} < 0$) than in accumulating ($\delta\rho_{EE} > 0$). As a consequence, the cumulative distribution functions (CDF) of $\delta\rho_{EE}$ in **Fig.7(a,b)** at $\phi = 0.671$ and $0.622$ have a value lower than 0.5 at $\delta\rho_{EE} \to 0$, regardless of being fully frictional, free-slipping or in quaking regimes. Moreover, **Fig.7(a)** highlights the fact that the quaking states spend a substantial fraction of time (a few percent, by reading the CDF value at $\delta\rho_{EE}/\overline{\rho_{EE}} = -0.1$ ) on very steep descending of $\rho_{EE}$, while that for those non-quaking states is well below 0.001 up to the same value of $\delta\rho_{EE}/\overline{\rho_{EE}}$. For the quaking states, the CDF also exhibits a plateau in $-0.1 < \delta\rho_{EE}/\overline{\rho_{EE}} < 0.001$, while in the same interval the non-quaking states incrementally accumulate a considerable amount of statistics (up to ~20%). Meanwhile, **Fig.7(b)** shows a visible but relatively subtle distinction between the Type-I and Type-II states: The curve for $S_\ell \sim 0.1$ appears to be a hybrid from the portions of $S_\ell \sim 0.01$ and $S_\ell \sim 1$, presumably as a consequence of being the "mixed state" — see supporting data in Supplemental Material for further discussions.

In **Fig.7(c),** we select four states along the path $\phi = 0.622$ with different values of $S_\ell$ and present the CDFs in linear scales. The plots reveal the development of an intriguing symmetry: The CDFs go across *precisely* 0.5 at $\delta\rho_{EE} = 0$ for the states with $S_\ell \geq 10$, indicating that the system spends *equal amounts of time in releasing and accumulating the elastic energy*. The effects of the force network, if any, become insignificant such that they no longer produce a detectable bias between the positive and negative branches of $\delta\rho_{EE}$.We regard it as a signature that the packing has been *fully disintegrated* due to combination of the lack of friction and insufficient packing density. The emergence of such statistical symmetry may also be anticipated for the "gaseous" state [1] of granular flow at much lower densities that are beyond the scope of our current studies, where all contacts are expected to be extremely short-lived (much below the $\delta t$ chosen for the statistical analysis here). In supporting data [Fig.S4 in SM], we provide representative values of the CDF showing how the statistical distribution of energy fluctuations loses its bias between positive and negative branches and becomes symmetric. Recalling Fig.3(a), it is based on those data that we draw the boundary for a transition toward "fully unjammed" states.

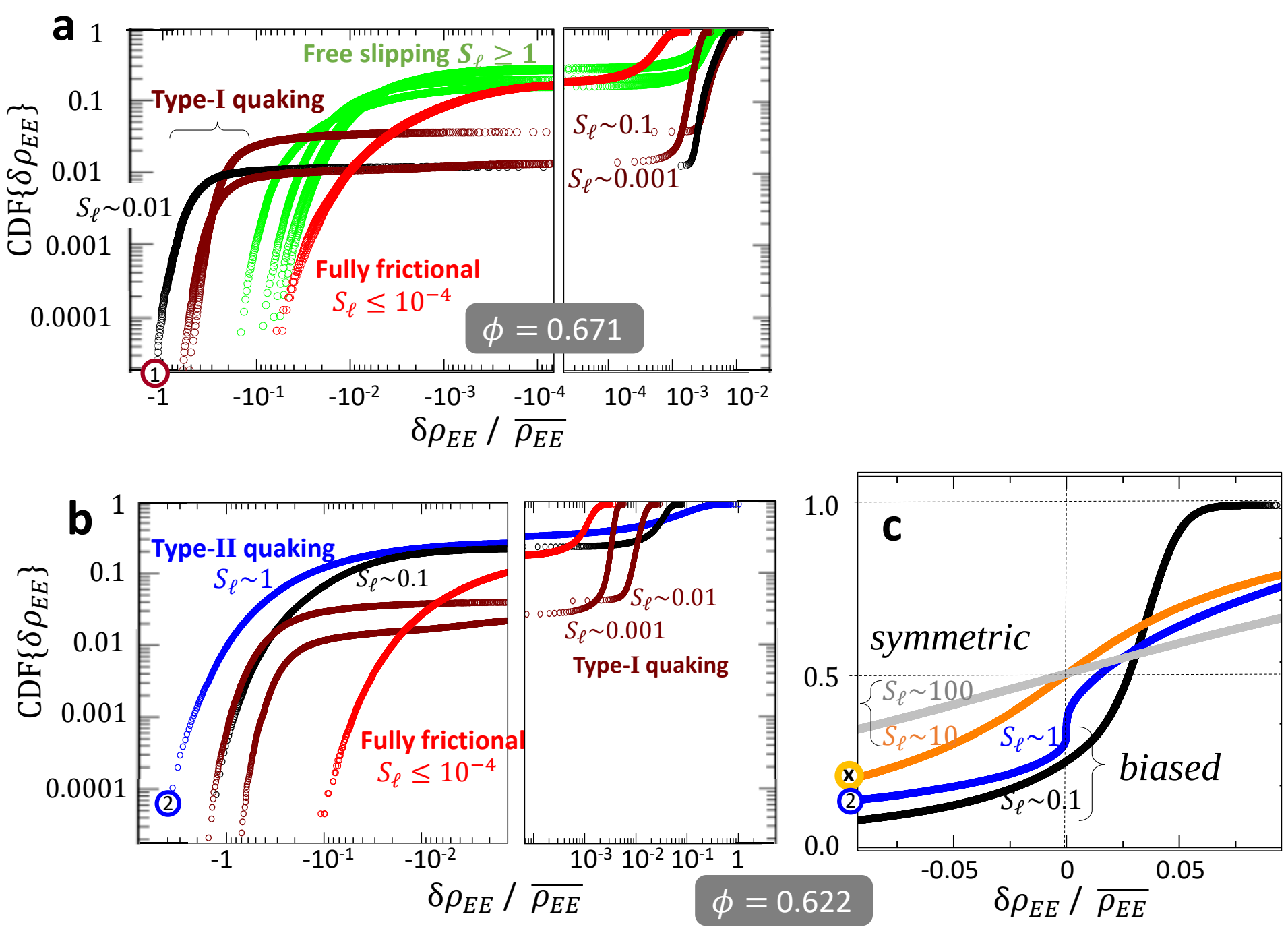

**FIG.7** – Cumulative distribution functions (CDFs) for $\delta\rho_{EE}/\overline{\rho_{EE}}$. **(a,b)** CDFs in logarithmic scales for the states with different values of $S_\ell$, along the two paths $\phi = 0.671$ and $\phi = 0.622$, respectively. **(c)** CDFs for $S_\ell$~$0.1, 1, 10$ and $100$ at $\phi = 0.622$, in linear scales.

## V. STATE DIAGRAM AND "SHEAR UNJAMMING"

**Figure 8** are schematic diagrams for a comparison between the states and transitions covered by the classic picture, that our data shown in Fig.2 with CH model are consistent with, and by the new perception obtained by implementing the SH model. Here, we use the packing fraction $\phi$ as the vertical axis, as it is the control parameter shared by both models. The left panel summarizes what have been well recognized in the research community. For instance, the benchmark work by Leo Silbert has established the "critical packing density" $\phi_C$ as a function of the friction coefficient $\mu_0$ [16]. Above $\phi_C$, the particles are considered "jammed" but are still able to move in response to shearing, because they are soft spheres. The result is insensitive to the shear rate as long as the system stays in the low-inertia limit. Over the past decade, a consensus in the community has been that the value of $\phi_C$ varies from $\phi_{RCP}$ ($\approx 0.64$ for frictionless spheres) to $\phi_{RLP}$ ($\approx 0.55$ at the frictional extreme) [16,17], and that a subtle scenario of jamming depending on whether the packing is being sheared or not is expected to occur between $\phi_{RCP}$ and $\phi_{RLP}$. The latter is known as "Jamming by Shear" [41] with many follow-up works reviewed in Ref.[41,44]. However, it should be noted that these perceptions have been developed in the context of a fixed friction coefficient.

Our numerical results using the SH model has shown, on the other hand, that the jamming-unjamming transition depends not only on $\phi$ and the action of shearing, but also on the dimensionless shear rate $S_\ell$. The underlying reasons are that (1) increasing $S_\ell$ leads to a decrease in the average friction, which subsequently raises the isostatic point, as discussed in Section III, and that (2) at packing fractions lower than $\phi_{RCP}$, increasing $S_\ell$ leads to the decrease of $\overline{n_C}$, as shown by the $\overline{n_C}$-$S_\ell$ diagram in Fig.3(a). As a result, the increase of $S_\ell$ eventually makes $\overline{n_C}$ go below the isostatic point so that the stability of the packing can no longer be maintained. Here, the right panel of Fig.8 recaptures the two types of quaking instability as described in Section II and III. In particular, between $\phi_{RCP}$ and a lower bound $\phi_0$ (slightly larger than $\phi_{RLP}$), the dominant quaking events can be either Type-I or Type-II --- depending upon the value of $S_\ell$. Given the bursts of displacements and the sudden drops of elastic energy of these quaking events, we interpret the granular system as being *intermittently unjammed* in the presence of such intermittency. In discrete time intervals, the granular packing are temporarily disintegrated. While such time intervals can be short and hard to measure in a state dominated by Type-I quaking, they are clearly detectable for a packing going through Type-II events, as we recall the close-up view of $\rho_{EE}(t)$ in Fig.6(d). Besides, at these intermediate values of $\phi$, increasing $S_\ell$ can ultimately render the packing *permanently disintegrated*, as is discussed with the emergence of a statistical symmetry in $\delta\rho_{EE}$ (Fig.7c, Section IV). This is marked by the yellow region labeled as "fully unjammed" on the lower right of Fig.8.

The two scenarios of unjamming provide a vivid example that the granular system can be unjammed *dynamically*, that is, via the increase of driving rate $S_\ell$ but with the packing fraction remaining unchanged.

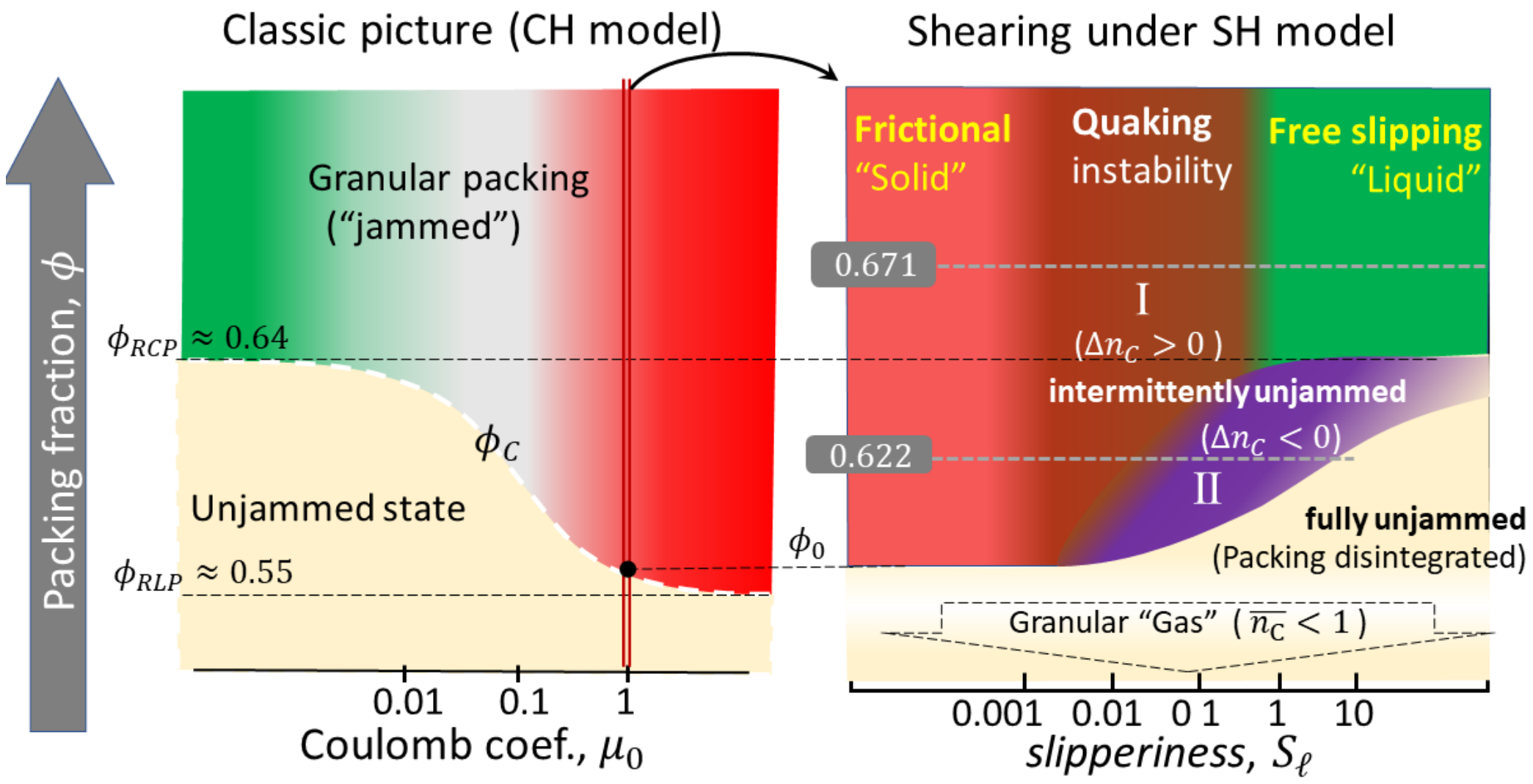

**FIG.8** –State Diagram with two scenarios of "shear unjamming" [Right, with $\mu_0(0)$=1 in SH model], compared to the classic understanding using a fixed friction [Left, CH model with a range of $\mu_0$]. On the right, we also denote the anticipated analogies to "three states of ordinary matter" by texts in quotation marks, in addition to the state transitions already described with Fig.3(a).

## VI. CONCLUDING REMARKS

In this work, we have shown that incorporating speed-dependent tribology at the level of individual particles has profound effects on the collective behavior of granular flow. We establish a state diagram spanned by two control parameters, the packing fraction ($\phi$) and a dimensionless shear rate ($S_\ell$, representing the *slipperiness* of the system). Our numerical studies successfully reproduce the rate-dependent intermittency discovered in previous experiments, and provide further predictions beyond existing laboratory observations. Notably, by studying the change in the mean contact number $n_C$ around particles, we have identified two types of intermittency. The first intermittency is the Type-I quaking characterized by a drastic increase in $n_C$. The second one is the Type-II quaking accompanied by a sudden decrease in $n_C$. For packing fractions higher than $\phi_{RCP}$, the events of quaking are exclusively of Type-I. In the range between $\phi_{RCP}$ and $\phi_{RLP}$, the dominant type of quaking depends further on the dimensionless shear rate $S_\ell$.These results unveils two distinct microstructural pathways underlying the quaking events, and elucidate the origin of intermittency that distinguishes granular flows from ordinary fluids.

We have also demonstrated that shearing can unjam a granular system dynamically. Based upon the analysis of fluctuations in elastic energy, the unjamming phenomena can be categorized into two scenarios: (1) intermittent unjamming in the presence of two types of quaking, and (2) permanent disintegration of the granular packing upon further increase of the dimensionless shear rate. These findings go beyond what is generally anticipated from the prevailing shear-jamming paradigm, which has been established in studying granular systems at a constant Coulomb friction. Given the fact that friction coefficients can vary quite substantially in the real world, our state diagram sheds light on revising the existing perception of the jamming-unjamming transition, for which granular flow has provided a vivid example.

# APPENDIX-A: FULL MODELS, PROTOCOLS, PARAMETERS AND CODING

Spherical particles of two sizes $d_A = 0.9d$ and $d_B = 1.1d$ are bounded by two walls formed by square array of particles with lattice spacing $d$, separated by a fixed distance $Z_0$. Shearing is applied by moving the two walls at a constant speed of $U/2$ in opposing directions along the $x$ direction. The box of simulation is periodic in $x$ and $y$, with all particles obeying Newtonian mechanics. The force between all pairs of particles follows a history-dependent, Cundall-Strack-like model implemented in LAMMPS [44–47]: For *Particle i* and *j* with a radial overlap $\vec{\delta}_{ji}$ and $R_{ij} \equiv \left(R_i^{-1} + R_j^{-1}\right)^{-1}$, the *contact* generates a normal force $\vec{f}_{ji}^{N} \equiv |\vec{\delta}_{ji}|^{0.5} \cdot R_{ij}^{0.5} \cdot K_N \cdot (\vec{\delta}_{ji} + \tau \cdot d\vec{\delta}_{ji}/dt)$ whose primary contribution is a Hertzian law, and a tangential force that is conditional on both the normal force and the accumulation of tangential displacement: $\vec{f}_{ji}^{T} \equiv |\vec{\delta}_{ji}|^{0.5} \cdot R_{ij}^{0.5} \cdot K_T \cdot \overrightarrow{\Delta S}_{ji}^{T}$ if such force has a magnitude smaller than $\mu \left| \vec{f}_{ji}^{N} \right|$, or simply $\mu \cdot \left| \vec{f}_{ji}^{N} \right| \cdot \overrightarrow{\Delta S}_{ji}^{T} / |\overrightarrow{\Delta S}_{ji}^{T}|$ if larger. Here, the vector $\overrightarrow{\Delta S}_{ji}^{T}$ stands for the lateral component of the relative boundary motion $\overrightarrow{\Delta S}_{ji}$ with respect to $\vec{\delta}_{ji}$, integrated since the start of a contact.

The key change we have made to the classic picture, which we denote as Coulomb-Hertzian (CH) model is to make $\mu$ dependent on the time derivative of $\overrightarrow{\Delta S}_{ji}^{T}$. This introduces a dependence on the sliding speed, as described in Fig.1(a). Our modification is denoted as Stribeck-Hertz (SH) model. In either case, the elastic energy stored in a contacting pair is determined by $E_N = R_{ij}^{0.5} \cdot K_N \cdot \left|\vec{\delta}_{ji}\right|^{2.5}/2.5$.

***Protocols*** –As shown by an animation available in Supplemental Material, particles are first created in an oversized space (with no overlap) before being compressed to reach the desired packing fraction $\phi$. The value of $\phi$ for each state under investigation is controlled by setting the separation distance $Z_0$ with the number of particles being fixed. Cyclic compressions and shearing, in combination of the use of bi-dispersed particles, generates various disordered configurations for independent runs. In the final stages, the simulation is run with an integrating time step $dt_\epsilon$ equal to $10^{-6}d/U$. To ensure reaching a steady state, a shear strain $\geq 30$ is accumulated for each run at a fixed value of $Z_0$, The time average of every physical quantity is then computed over a shear strain $\sim O(1)$ at the end of the steady shearing to establish the state diagram.

***Dynamical regime and parameters*** – The choice of $U$= 0.1cm/s, $d$=1cm, mass $m_A = m_B$ = 1g, elastic modulus $K = K_n = K_T$ = 1.5MPa, the range of $Z_0$ and the packing fraction $\phi$ in this study ensures that the density of *kinetic energy* (~mass density $* d^2\dot{\gamma}^2$) is much smaller than that of elastic energy, which gives *Inertia Number* [48] << $10^{-2}$ to make sure that our "granular packing" is well within the low-inertia limit --- see supporting data [Fig.S1 in SM]. Such condition is held by all simulations reported here with either the SH model or the CH model. We have also checked that all results are not sensitive to the changes of the damping constant $\tau$ in the force law, for a range $10^{-2}$~$10^{-4}$s. The numbers of the two kinds of particles in our standard simulations are $N_A$= 576 and $N_B$= 1091, in a box of $12d \times 8d$ on the *x-y* plane with the height $Z_0$ being $14 \sim 20d$.

Technical notes, including key changes made to the existing LAMMPS codes (in CH model) for implementing our SH simulations, pending modifications on LAMMPS for public use, and guides to representative datasets, are offered in **Supplemental Material.** 3D graphics and animations of our results are generated by Ovito [49,50].

We anticipate our numerical experiments to be extended to go beyond the low-inertia limit and incorporate the entire Stribeck curve --- see supporting data [Fig.S5B in SM]. We are also aware of a few recent works such as one by Bonn and coworkers [51] in which both the tribology and the rejuvenation dynamics are considered. Nevertheless, in our current studies, we intend to use a minimal set of assumptions in order to pinpoint the role of tribology by itself.

# References

[1] I. Goldhirsch, *Rapid Granular Flows*, Annu. Rev. Fluid Mech. **35**, 267 (2003).

[2] T. S. Komatsu, S. Inagaki, N. Nakagawa, and S. Nasuno, *Creep Motion in a Granular Pile Exhibiting Steady Surface Flow*, Phys. Rev. Lett. **86**, 1757 (2001).

[3] J.-C. Tsai and J. P. Gollub, *Slowly Sheared Dense Granular Flows: Crystallization and Nonunique Final States*, Phys. Rev. E **70**, 031303 (2004).

[4] J. A. Dijksman and M. van Hecke, *Granular Flows in Split Bottom Geometries*, AIP Conf. Proc. **1145**, 56 (2009).

[5] K. E. Daniels, J. E. Kollmer, and J. G. Puckett, *Photoelastic Force Measurements in Granular Materials*, Rev. Sci. Instrum. **88**, 051808 (2017).

[6] D. M. Mueth, G. F. Debregeas, G. S. Karczmar, P. J. Eng, S. R. Nagel, and H. M. Jaeger, *Signatures of Granular Microstructure in Dense Shear Flows*, Nature **406**, 6794 (2000).

[7] C. M. Boyce, N. P. Rice, A. Ozel, J. F. Davidson, A. J. Sederman, L. F. Gladden, S. Sundaresan, J. S. Dennis, and D. J. Holland, *Magnetic Resonance Characterization of Coupled Gas and Particle Dynamics in a Bubbling Fluidized Bed*, Phys. Rev. Fluids **1**, 074201 (2016).

[8] J.-C. Tsai, G. A. Voth, and J. P. Gollub, *Internal Granular Dynamics, Shear-Induced Crystallization, and Compaction Steps*, Phys. Rev. Lett. **91**, 064301 (2003).

[9] Y. Yuan et al., *Experimental Test of the Edwards Volume Ensemble for Tapped Granular Packings*, Phys. Rev. Lett. **127**, 018002 (2021).

[10] J. A. Dijksman, F. Rietz, K. A. Lőrincz, M. van Hecke, and W. Losert, *Invited Article: Refractive Index Matched Scanning of Dense Granular Materials*, Rev. Sci. Instrum. **83**, 011301 (2012).

[11] B. Miller, C. O'Hern, and R. P. Behringer, *Stress Fluctuations for Continuously Sheared Granular Materials*, Phys. Rev. Lett. **77**, 3110 (1996).

[12] D. Lootens, H. Van Damme, and P. Hébraud, *Giant Stress Fluctuations at the Jamming Transition*, Phys. Rev. Lett. **90**, 178301 (2003).

[13] J.-C. (JC) Tsai, G.-H. Huang, and C.-E. Tsai, *Signature of Transition between Granular Solid and Fluid: Rate-Dependent Stick Slips in Steady Shearing*, Phys. Rev. Lett. **126**, 128001 (2021).

[14] F. Boyer, É. Guazzelli, and O. Pouliquen, *Unifying Suspension and Granular Rheology*, Phys. Rev. Lett. **107**, 188301 (2011).

[15] É. Guazzelli and O. Pouliquen, *Rheology of Dense Granular Suspensions*, J. Fluid Mech. **852**, P1 (2018).

[16] L. E. Silbert, *Jamming of Frictional Spheres and Random Loose Packing*, Soft Matter **6**, 2918 (2010).

[17] C. Song, P. Wang, and H. A. Makse, *A Phase Diagram for Jammed Matter*, Nature **453**, 629 (2008).

[18] S. Papanikolaou, C. S. O'Hern, and M. D. Shattuck, *Isostaticity at Frictional Jamming*, Phys. Rev. Lett. **110**, 198002 (2013).

[19] J. R. Royer and P. M. Chaikin, *Precisely Cyclic Sand: Self-Organization of Periodically Sheared Frictional Grains*, Proc. Natl. Acad. Sci. **112**, 49 (2015).

[20] R. Mari, R. Seto, J. F. Morris, and M. M. Denn, *Shear Thickening, Frictionless and Frictional Rheologies in Non-Brownian Suspensions*, J. Rheol. **58**, 1693 (2014).
[21] J. F. Morris, *Shear Thickening of Concentrated Suspensions: Recent Developments and Relation to Other Phenomena*, Annu. Rev. Fluid Mech. **52**, 121 (2020).
[22] S. Jamali and J. F. Brady, *Alternative Frictional Model for Discontinuous Shear Thickening of Dense Suspensions: Hydrodynamics*, Phys. Rev. Lett. **123**, 138002 (2019).
[23] I. S. Aranson, L. S. Tsimring, F. Malloggi, and E. Clément, *Nonlocal Rheological Properties of Granular Flows near a Jamming Limit*, Phys. Rev. E **78**, 031303 (2008).
[24] D. L. Henann and K. Kamrin, *A Predictive, Size-Dependent Continuum Model for Dense Granular Flows*, Proc. Natl. Acad. Sci. **110**, 6730 (2013).
[25] Z. Tang, T. A. Brzinski, M. Shearer, and K. E. Daniels, *Nonlocal Rheology of Dense Granular Flow in Annular Shear Experiments*, Soft Matter **14**, 3040 (2018).
[26] E. E. Brodsky et al., *The State of Stress on the Fault Before, During, and After a Major Earthquake*, Annu. Rev. Earth Planet. Sci. **48**, 49 (2020).
[27] G. Di Toro, R. Han, T. Hirose, N. De Paola, S. Nielsen, K. Mizoguchi, F. Ferri, M. Cocco, and T. Shimamoto, *Fault Lubrication during Earthquakes*, Nature **471**, 7339 (2011).
[28] K.-F. Ma, E. E. Brodsky, J. Mori, C. Ji, T.-R. A. Song, and H. Kanamori, *Evidence for Fault Lubrication during the 1999 Chi-Chi, Taiwan, Earthquake (Mw7.6)*, Geophys. Res. Lett. **30**, (2003).
[29] K. Aki and P. G. Richards, *Quantitative Seismology* (University Science Books, 2002).
[30] P. M. Shearer, *Introduction to Seismology* (Cambridge University Press, 2019).
[31] G. Di Toro, T. Hirose, S. Nielsen, G. Pennacchioni, and T. Shimamoto, *Natural and Experimental Evidence of Melt Lubrication of Faults During Earthquakes*, Science **311**, 647 (2006).
[32] B. Ferdowsi and A. M. Rubin, *A Granular Physics-Based View of Fault Friction Experiments*, J. Geophys. Res. Solid Earth **125**, e2019JB019016 (2020).
[33] L.-W. Kuo et al., *Frictional Properties of the Longmenshan Fault Belt Gouges From WFSD-3 and Implications for Earthquake Rupture Propagation*, J. Geophys. Res. Solid Earth **127**, e2022JB024081 (2022).
[34] J. Williams, *Engineering Tribology* (Cambridge University Press, Cambridge, 2005).
[35] B. Jacobson, *The Stribeck Memorial Lecture*, Tribol. Int. **36**, 781 (2003).
[36] M. Workamp and J. A. Dijksman, *Contact Tribology Also Affects the Slow Flow Behavior of Granular Emulsions*, J. Rheol. **63**, 275 (2019).
[37] C.-E. Tsai, Master Thesis, Physics Dept., National Central University, Taoyan, Taiwan, 2022.
[38] C.-E. Tsai and J.-C. Tsai, *Force between Contacting PDMS Surfaces upon Steady Sliding: Speed Dependence and Fluctuations*, http://arxiv.org/abs/2010.07907.
[39] K. E. Daniels and N. W. Hayman, *Force Chains in Seismogenic Faults Visualized with Photoelastic Granular Shear Experiments*, J. Geophys. Res. Solid Earth **113**, (2008).
[40] M. E. Cates, J. P. Wittmer, J.-P. Bouchaud, and P. Claudin, *Jamming, Force Chains, and Fragile Matter*, Phys. Rev. Lett. **81**, 1841 (1998).
[41] D. Bi, J. Zhang, B. Chakraborty, and R. P. Behringer, *Jamming by Shear*, Nature **480**, 355 (2011).

[42] R. P. Behringer and B. Chakraborty, *The Physics of Jamming for Granular Materials: A Review*, Rep. Prog. Phys. **82**, 012601 (2018).

[43] Y. Zhao, Y. Zhao, D. Wang, H. Zheng, B. Chakraborty, and J. E. S. Socolar, *Ultrastable Shear-Jammed Granular Material*, Phys. Rev. X **12**, 031021 (2022).

[44] *Gran/Hertz/History in LAMMPS Documentation*, https://lammps.sandia.gov/.

[45] A. P. Thompson et al., *LAMMPS - a Flexible Simulation Tool for Particle-Based Materials Modeling at the Atomic, Meso, and Continuum Scales*, Comput. Phys. Commun. **271**, 108171 (2022).

[46] L. E. Silbert, D. Ertaş, G. S. Grest, T. C. Halsey, D. Levine, and S. J. Plimpton, *Granular Flow down an Inclined Plane: Bagnold Scaling and Rheology*, Phys. Rev. E **64**, 051302 (2001).

[47] H. P. Zhang and H. A. Makse, *Jamming Transition in Emulsions and Granular Materials*, Phys. Rev. E **72**, 011301 (2005).

[48] GDR MiDi, *On Dense Granular Flows*, Eur. Phys. J. E **14**, 341 (2004).

[49] *OVITO*, http://www.ovito.org/.

[50] A. Stukowski, *Visualization and Analysis of Atomistic Simulation Data with OVITO–the Open Visualization Tool*, Model. Simul. Mater. Sci. Eng. **18**, 015012 (2009).

[51] K. Farain and D. Bonn, *Quantitative Understanding of the Onset of Dense Granular Flows*, Phys. Rev. Lett. **130**, 108201 (2023).